\documentclass{aa}

\usepackage{upgreek}
\usepackage{txfonts}
\usepackage{graphicx}
\usepackage{xcolor}
\usepackage[colorlinks=true, linkcolor=blue, citecolor=blue, urlcolor=blue]{hyperref}
\usepackage{natbib}

\bibpunct{(}{)}{;}{a}{}{,} 

\newcommand{\txd}{{\text{d}}}
\newcommand{\calE}{{\cal{E}}}

\begin{document}

\title{The Einasto model for dark matter haloes}

\author{Maarten Baes}

\institute{
Sterrenkundig Observatorium, Universiteit Gent, Krijgslaan 281 S9, 9000 Gent, Belgium \\ email: {\tt{maarten.baes@ugent.be}}
}

\date{Received 22 July 2022 / Accepted 5 September 2022}

\abstract{The Einasto model has become one of the most popular models for describing the density profile of dark matter haloes. There have been relatively few comprehensive studies on the dynamical structure of the Einasto model, mainly because only a limited number of properties can be calculated analytically.}
{We want to systematically investigate the photometric and dynamical structure of the family of Einasto models over the entire model parameter space.}
{We used the {\tt{SpheCow}} code to explore the properties of the Einasto model. We systematically investigated how the most important properties change as a function of the Einasto index $n$. We considered both isotropic models and radially anisotropic models with an Osipkov-Merritt orbital structure.}
{We find that all Einasto models with $n<\tfrac12$ have a formal isotropic or Osipkov-Merritt distribution function that is negative in parts of phase space, and hence cannot be supported by such orbital structures. On the other hand, all models with larger values of $n$ can be supported by an isotropic orbital structure, or by an Osipkov-Merritt anisotropy, as long as the anisotropy radius is larger than a critical value. This critical anisotropy radius is a decreasing function of $n$, indicating that less centrally concentrated models allow for a larger degree of radial anisotropy.}
{Studies of the structure and dynamics of models for galaxies and dark matter haloes should not be restricted to completely analytical models. Numerical codes such as {\tt{SpheCow}} can help open up the range of models that are systematically investigated. This applies to the Einasto model discussed here, but also to other proposed models for dark matter haloes, including different extensions to the Einasto model.}


\maketitle

\section{Introduction}

According to the $\Lambda$ Cold Dark Matter model, the matter budget of the Universe is dominated by dark matter. Based on recent Planck-based cosmological model parameter estimates, about 85\% of all matter in the Universe consists of dark matter \citep{2020A&A...641A...6P}. Dark matter haloes, the highest density structures within the cosmic web, contain a large fraction of the dark matter content and most of the baryonic matter in the Universe. The characterisation of dark matter haloes is important, and this is usually done by means of their spherically averaged density profiles. 

The most popular model for describing the density profile of dark matter haloes is probably the Navarro, Frenk \& White (NFW) model \citep{1997ApJ...490..493N}. This model is characterised by an $r^{-1}$ power-law behaviour at small radii and an $r^{-3}$ power-law slope at large radii. It belongs to a larger family of double power-law models, characterised by different asymptotic power laws, $r^{-\gamma}$ and $r^{-\beta}$, at small and large radii, respectively. Many different members of this large family have been proposed as models for dark matter haloes \citep[e.g.][]{1999MNRAS.310.1147M, 2000ApJ...529L..69J, 2000ApJ...544..616G, 2014MNRAS.441.2986D, 2014MNRAS.443.3712H, 2015ApJ...800...15H, 2017MNRAS.468.1005D}. The dynamical and lensing properties of this set of double power-law models have been explored in quite some detail \citep{1996MNRAS.278..488Z, 2001MNRAS.321..155L, 2005MNRAS.360..492E, 2020MNRAS.499.2912F, 2021MNRAS.503.2955B}.

A different model in which the logarithmic density slope rather than the density itself shows a power-law behaviour was proposed by \citet{2004MNRAS.349.1039N} and \citet{2005ApJ...624L..85M}, arguing that this model provided a better fit to the density profile of high-resolution N-body dark matter haloes than the NFW model. Models with a power-law logarithmic density slope were originally introduced by \citet{1965TrAlm...5...87E}, and used to model the stellar distribution in nearby galaxies such as M31, M32, and M87 and in the Milky Way \citep{1969Afz.....5..137E, 1974smws.conf..291E}. The model is currently generally known as the Einasto model. It has been used extensively to describe simulated dark matter haloes \citep[e.g.][]{2006AJ....132.2685M, 2008MNRAS.387..536G, 2008MNRAS.388....2H, 2008MNRAS.391.1685S, 2010MNRAS.402...21N, 2014MNRAS.439..300L, 2016MNRAS.457.4340K, 2020MNRAS.499.2426F, 2020Natur.585...39W} and to fit galaxy rotation curves \citep[e.g.][]{2011AJ....142..109C, 2019MNRAS.482.5106L, 2019A&A...623A.123G}.

The properties of the Einasto model have been investigated by a number of authors \citep{2005MNRAS.362...95M, 2005MNRAS.358.1325C, 2006AJ....132.2685M, 2010A&A...520A..30M, 2010MNRAS.405..340D, 2012A&A...540A..70R, 2012A&A...546A..32R, 2021MNRAS.504.4583D}, but not to the same degree as other models such as the double power-law model. One reason is that relatively few properties of the Einasto model can be calculated analytically. The density profile is a simple analytical function, but even simple derived quantities such as the mass profile or the gravitational potential cannot be expressed without the use of special transcendental functions. Projected and lensing properties can only formally be expressed using very special functions such as the Fox H function. There is no hope to express more intricate dynamical properties such as the distribution function or the differential energy distribution as a closed expression. 

The most comprehensive studies on the family of Einasto models are those by \citet{2005MNRAS.358.1325C} and \citet{2012A&A...540A..70R}. In the former paper, analytical expressions for the basic properties are derived and the observed on-sky properties and the distribution function and differential energy distribution for an isotropic and Osipkov-Merritt orbital structure are calculated numerically and discussed. The latter paper focuses primarily on the analytical calculations of the surface density and the lensing properties and on a comparison between the Einasto and S\'ersic models.

There are several reasons why these studies should be refined and extended. First of all, they consider only a limited range in the Einasto parameter, $n$, focusing on large values. Inspired by our recent work on the family of S\'ersic models \citep{2019A&A...626A.110B}, we argue that models with smaller values of $n$, which correspond to a sharper transition between the inner and outer part of the model, can be significantly different and deserve the necessary attention. Secondly, a number of results in the \citet{2005MNRAS.358.1325C} study are intriguing. For example, the authors found that the shape of the differential energy distribution of the family of Einasto models depends sensitively on $n$. In the range of small binding energies, it seems to converge to zero for smaller values of $n$, as expected. For the highest values of $n$ they considered, $n\approx10$, the differential energy distribution seems to diverge at small binding energies. The authors suggest that this unexpected behaviour could be the result of numerical problems in their integrations. Finally, \citet{2005MNRAS.358.1325C} discuss anisotropic models with an Osipkov-Merritt orbital structure, but do not discuss the consistency limits, whereas this is a crucial characteristic for this kind of dynamical model \citep[e.g.][]{1979PAZh....5...77O, 1985AJ.....90.1027M, 1995MNRAS.276.1131C, 1997A&A...321..724C, 2002A&A...393..485B}.  

We have recently released {\tt{SpheCow}}, a software tool designed to numerically explore the dynamical structure of any spherical model defined by an analytical density profile or surface density profile \citep{2021A&A...652A..36B}. The code can be used to numerically explore the dynamical structure, under the assumption of an isotropic or an Osipkov-Merritt anisotropic orbital structure. {\tt{SpheCow}} contains readily usable implementations for many standard models, including the Plummer, Hernquist, NFW, S\'ersic, Nuker, and Zhao models. Applications of {\tt{SpheCow}} include a detailed analysis of the families of Nuker and S\'ersic models \citep{2020A&A...634A.109B, 2019A&A...626A.110B, 2019A&A...630A.113B}, a study of the consistency of the double and broken power-law models \citep{2021MNRAS.503.2955B}, and the presentation of two new families of models with sigmoid density slopes \citep{2021A&A...652A..36B}. 

In this paper we make use of the capabilities of {\tt{SpheCow}} to fully explore the photometric and dynamical structure of the family of Einasto models. Throughout this paper we focus on the results and the interpretation of the findings, rather than on the derivation of the relevant formulae. For readers less familiar with the construction of spherical dynamical models and the formulae involved, we refer to the {\tt{SpheCow}} presentation paper \citep{2021A&A...652A..36B} or to standard stellar dynamics textbooks \citep[e.g.][]{2008gady.book.....B, 2021isd..book.....C} for more information.

This paper is organised as follows. In Sect.~{\ref{Einastomodel.sec}} we introduce the Einasto model and focus on the dimensionless parameter $d$ that appears in the formal expression of the density. In Sect.~{\ref{Einasto-basic.sec}} we discuss the basic properties for the full range of the Einasto parameter, $n$. In Sect.~{\ref{isotropic.sec}} we investigate the dynamical properties of the isotropic Einasto models, focusing on the velocity dispersions, the distribution function, and the differential energy distribution. In Sect.~{\ref{OM.sec}} we discuss anisotropic Einasto models with an Osipkov-Merritt orbital structure and we particularly investigate the consistency for different Einasto parameters and anisotropy radii. In Sect.~{\ref{Discussion.sec}} we discuss and summarise our findings.

\section{The Einasto model}
\label{Einastomodel.sec}

\subsection{Definition of the model}

The Einasto model is characterised by the density profile
\begin{equation}
\rho(r) = \frac{d^{3n}}{4\pi\,n\,\Gamma(3n)}\,\frac{M}{r_{\text{h}}^3} \exp\left[-d\left(\frac{r}{r_{\text{h}}}\right)^{1/n}\right].
\label{rho}
\end{equation}
It contains three free parameters: the total mass, $M$, the half-mass radius, $r_{\text{h}}$, and the Einasto index, $n$, that characterises the concentration of the mass distribution. While $M$ and $r_{\text{h}}$ can be regarded as scaling parameters, the Einasto index is the only shape parameter that determines the actual shape of the density profile. 

We note that there are different parameterisations in the literature for the Einasto model, with different free parameters. In the context of dark matter haloes, a popular representation is
\begin{equation}
\rho(r) = \rho_{-2} \exp\left\{ -2n \left[ \left(\frac{r}{r_{-2}}\right)^{1/n}-1 \right] \right\},
\end{equation}
where $r_{-2}$ and $\rho_{-2}$ represent the radius and the density at which the logarithmic density slope is equal to $-2$. For other conventions and the conversion between them, we refer to \citet{2012A&A...540A..70R}.

\subsection{The dimensionless parameter $d$}

\begin{figure}
\includegraphics[width=0.86\columnwidth]{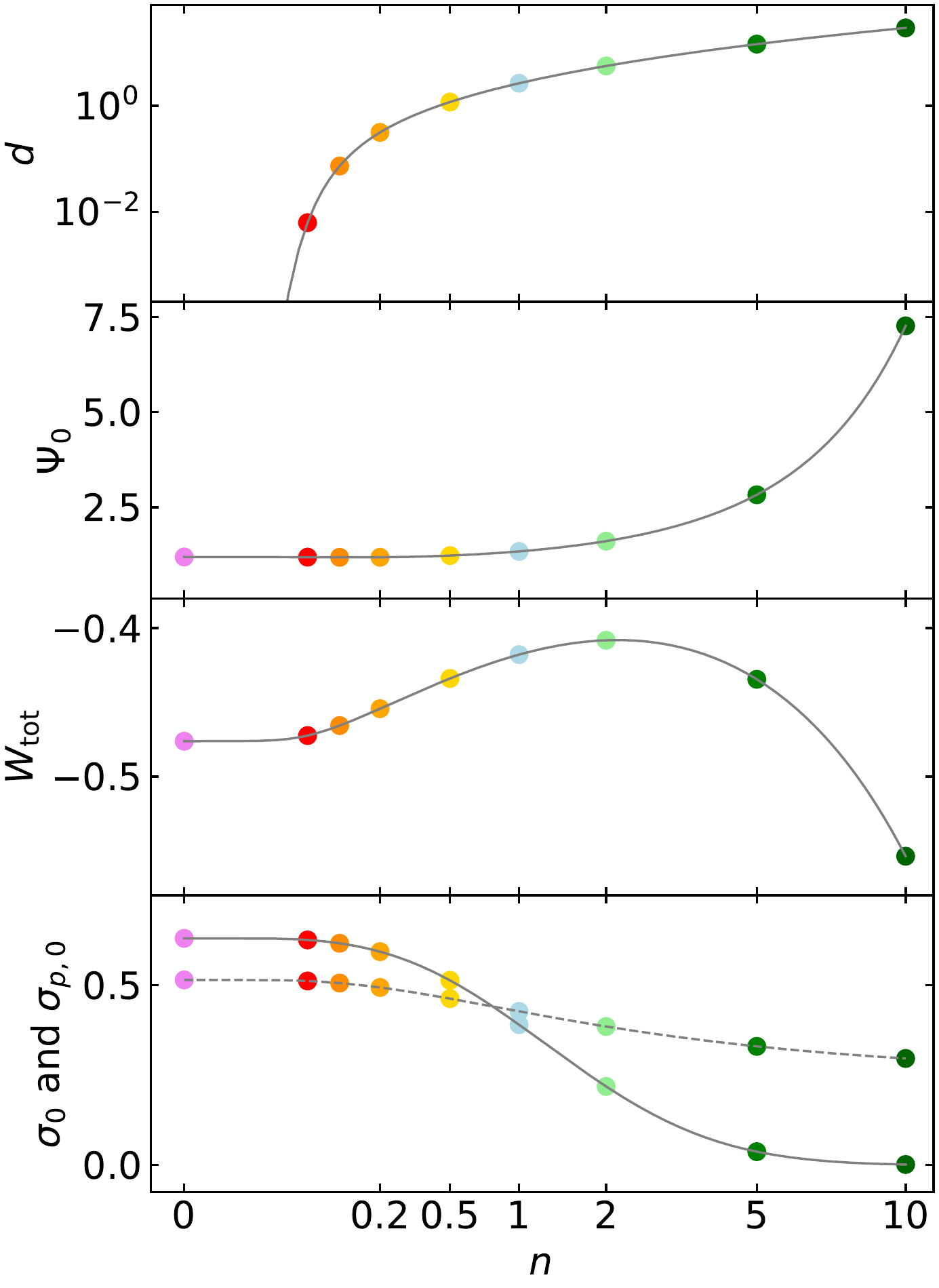}%
\caption{Dependence of some global and dynamical properties for the family of Einasto model, explicitly as a function of the Einasto index and for fixed total mass and half-mass radius. From top to bottom this figure shows the dimensionless quantity $d$, the depth of the potential well, the total potential energy, and the central intrinsic (solid line) and line-of-sight (dash line) velocity dispersion. The latter two quantities are correspond to an isotropic dynamical structure. The solid lines correspond to analytical formulae or numerical results calculated outside {\tt{SpheCow}}, the coloured dots are numerical results calculated with {\tt{SpheCow}}. The colours of the dots correspond to the models shown in Fig.~{\ref{Einasto-basic.fig}}. }
\label{Einasto-central.fig}
\end{figure}

Apart from the free parameters $M$, $r_{\text{h}}$, and $n$, expression~(\ref{rho}) contains the parameter $d$. This is not a free parameter but a dimensionless constant that depends on $n$ and that is introduced to guarantee that $r_{\text{h}}$ is indeed the half-mass radius. Using the expression for the cumulative mass profile of the Einasto model \citep{2012A&A...540A..70R}, it is straightforward to show that $d$ can be found as the solution of the equation
\begin{equation}
\Gamma(3n,d) = \frac12\,\Gamma(3n),
\end{equation}
where $\Gamma(s)$ and $\Gamma(s,x)$ are the complete and the upper incomplete gamma functions, respectively. This equation can be solved numerically for any value of $n>0$.

Using an asymptotic expansion technique first explored by \citet{1999A&A...352..447C}, \citet{2012A&A...540A..70R} found the following asymptotic expansion for $d$, appropriate for the range $n>1$,
\begin{multline}
d = 3n-\frac13 + \frac{8}{1215\,n} + \frac{184}{229635\,n^2}
\\
+ \frac{1048}{31000725\,n^3} - \frac{17557576}{1242974068875\,n^4} + {\mathcal{O}}\left(\frac{1}{n^5}\right).
\label{d(n)}
\end{multline}
For small values of $n$, this expansion obviously does not hold and it seems an obvious option to use a power-series approximation for the small $n$ range. We can find inspiration from the study by \citet{2019A&A...626A.110B} for the equivalent parameter $b$ for the family of S\'ersic models. Using a similar argumentation and methodology, we find that $d$ tends to zero as $n$ as $n$ approaches zero, but that $d^n$ reaches a finite value in the limit $n\to0$,
\begin{equation}
\lim_{n\to0} d^n = \frac{1}{\sqrt[3]2}.
\end{equation}
We find that $d^n$ is very well fitted by a polynomial approximation over the interval $0\leqslant n \leqslant 1$,
\begin{equation}
d^n \approx \frac{1}{\sqrt[3]{2}} + \sum_{k=1}^4 a_k\,n^k
\end{equation}
with coefficients
\begin{gather}
a_1 = -0.4977745, \\
a_2 = 2.894682, \\
a_3 = -2.369477, \\
a_4 = 1.852409.
\end{gather}
The corresponding approximation formula for $d$,
\begin{equation}
d \approx \left(\frac{1}{\sqrt[3]{2}} + \sum_{k=1}^4 a_k\,n^k\right)^{1/n},
\end{equation}
is characterised by a relative RMS error of $4.5\times10^{-4}$ and an absolute RMS error of $4.6\times10^{-5}$ when averaged over the interval $0\leqslant n \leqslant 1$. The dependence of $d$ as a function of $n$ is shown in the top panel of Fig.~{\ref{Einasto-central.fig}}.

\section{Basic properties}
\label{Einasto-basic.sec}

\begin{figure*}
\hspace*{2em}%
\includegraphics[width=0.9\textwidth]{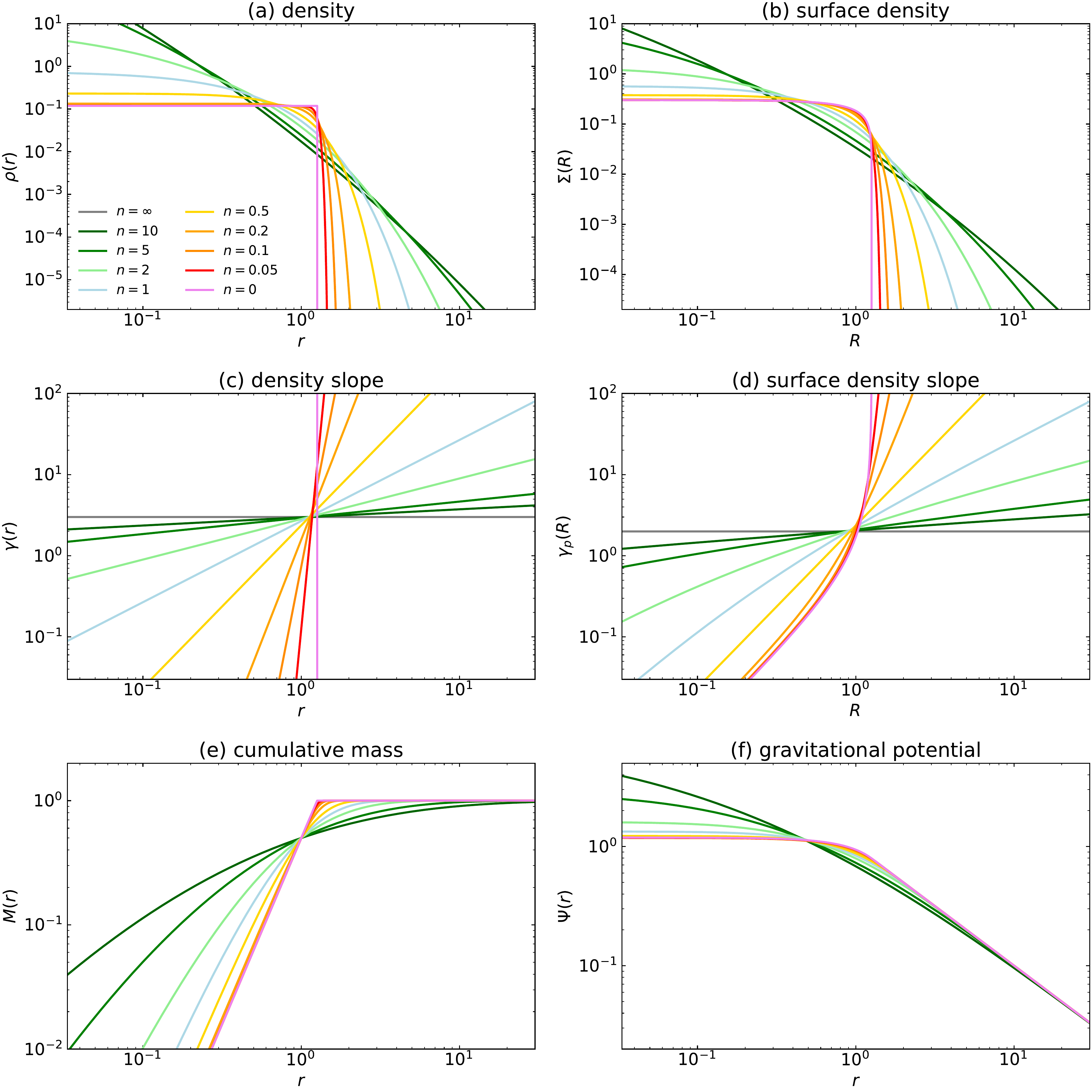}%
\caption{Basic properties of the family of Einasto models. We have set $M=r_{\text{h}}=1$. The different lines in each panel correspond to different values of $n$, as indicated in panel (a).}
\label{Einasto-basic.fig}
\end{figure*}

In Fig.~{\ref{Einasto-basic.fig}} we plot a number of basic properties of the family of Einasto models, as calculated with the {\tt{SpheCow}} code, for a range of values for the Einasto parameter $n$. 

\subsection{Density and surface density}
\label{density.sec}

Panels (a) and (b) show the density and the surface density, respectively. The surface density of the Einasto model cannot be calculated analytically in terms of elementary functions or even the regular special functions. It is possible to derive an expression in terms of the Fox $H$ function \citep{2012A&A...540A..70R} and various numerical approximations have been presented \citep{2010MNRAS.405..340D, 2021MNRAS.504.4583D}. Comparing the corresponding curves in both panel, it appears that, at first sight, the density profile and the surface density profile have a qualitatively similar behaviour. Both profiles are characterised by a finite value at the centre and in both cases, the profile has a smooth curvature (in log-log space) with a slope that gradually changes from the inner to the outer regions. For large values of $n$, the change is very smooth and the transition gets gradually sharper as $n$ decreases. 

The main difference between both profiles is more clearly visible when we compare the negative logarithmic slopes of both profiles, defined as
\begin{gather}
\gamma(r) = -\frac{\txd\log\rho}{\txd\log r}(r), \\
\gamma_{\text{p}}(R) = -\frac{\txd\log\Sigma}{\txd\log R}(R).
\end{gather}
These slope profiles are shown in the panels (c) and (d), respectively. The slope of the density profile is a pure power law over the entire radial range,
\begin{equation}
\gamma(r) = \frac{d}{n}\left(\frac{r}{r_{\text{h}}}\right)^{1/n},
\label{Einasto-gamma}
\end{equation}
which is exactly the defining property for the Einasto models \citep{2005MNRAS.358.1325C}. For the surface density profile, this is not the case: $\gamma_{\text{p}}(R)$ behaves as a power law at both small and large radii, but not over the entire radial range. 

These logarithmic slope plots are also useful to investigate the behaviour of the Einasto model in the limits of very large and very small Einasto indices. When $n$ becomes very large, $\gamma(r)$ becomes very flat, that is, with a slope that is almost uniform at all radii. Taking the limit $n\to\infty$ in expression~(\ref{Einasto-gamma}), we have, considering the expansion (\ref{d(n)}), $\lim_{n\to\infty} \gamma(r) = 3$. The limiting case of the Einasto models for $n\to\infty$ is hence a scale-free power-law model \citep{1994MNRAS.267..333E}, with $\rho(r)\propto r^{-3}$. The corresponding surface density profile is also a pure power law, $\Sigma(R)\propto R^{-2}$. Unfortunately, scale-free power-law models always have an infinite total mass, with the mass diverging in the centre if the power-law slope is larger than or equal to three, and at large radii if the slope is smaller than or equal to three. In the limiting case of the Einasto model, the mass profile hence diverges at both small and large radii. This model is represented by the grey line in panels (c) and (d).

In the other limiting case, corresponding to very small Einasto indices, we find a model with a constant density at small radii, an infinitely sharp break in the density profile, and a vanishing density beyond that point. We hence end up with a uniform density sphere with radius 
\begin{equation}
r_{\text{max}} = \sqrt[3]{2}\,r_{\text{h}}. 
\end{equation}
The uniform density sphere is one of the classical models used in the theory of dynamics \citep{1971Afz.....7..223B, 2021Ap.....64..219B, 1974SvA....17..460P, 1979PAZh....5...77O} and gravitational lensing studies \citep{1972MNRAS.158..233C, 2002MNRAS.337.1269W, Suyama_2005}. A detailed analysis of the dynamical properties of the uniform density sphere was recently presented by \citet{2022MNRAS.512.2266B}. The model is represented as the violet line in Fig.~{\ref{Einasto-basic.fig}}.

\begin{figure*}
\hspace*{2em}%
\includegraphics[width=0.9\textwidth]{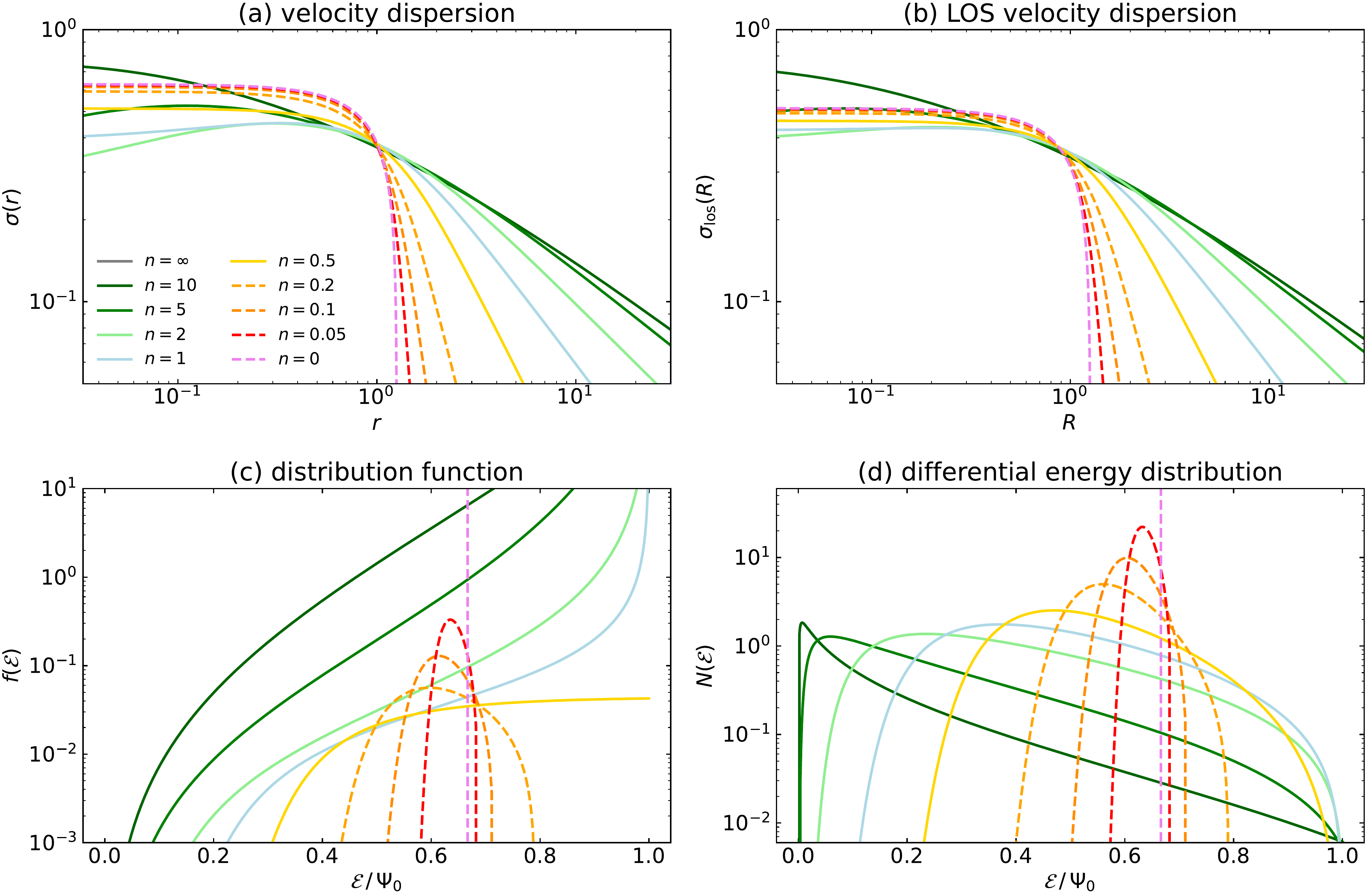}%
\caption{Dynamical properties of the family of Einasto models with an isotropic orbital structure. As in Fig.~{\ref{Einasto-basic.fig}} we have set $M=r_{\text{h}}=1$, and the different lines correspond to different values of $n$, as indicated in panel (a). Solid lines correspond to consistent isotropic models, dashed lines to unphysical dynamical models.}
\label{Einasto-isotropic.fig}
\end{figure*}

\subsection{Cumulative mass and potential}

Panels (e) and (f) show the cumulative mass profile and the gravitational potential. As demonstrated by \citet{2005MNRAS.358.1325C} and \citet{2012A&A...540A..70R}, all Einasto models have a finite mass and a finite potential well. The potential decreases parabolically at small radii, and in a Keplerian way at large radii. In these three panels we also over-plotted the profiles for the uniform density sphere \citep{2022MNRAS.512.2266B}. It is clear that the family of Einasto models nicely converges to the uniform density sphere as $n$ decreases.

In the second panel of Fig.~{\ref{Einasto-central.fig}} we explicitly show the depth of the potential well as a function of the Einasto index $n$, for fixed values of the total mass and the half-mass radius. The coloured dots correspond to the values calculated with {\tt{SpheCow}}, whereas the solid line represents the exact value~\citep{2012A&A...540A..70R},
\begin{equation}
\Psi_0 = \frac{GM}{r_{\text{h}}}\,\frac{1}{d^n}\, \frac{\Gamma(3n)}{\Gamma(2n)}.
\end{equation}
Interestingly, the depth of the potential well is not a monotonic function of $n$ for fixed values of the total mass and the half-mass radius: the minimum value is obtained for $n=0.145$.

The third panel of Fig.~{\ref{Einasto-central.fig}} shows the total potential energy as a function of $n$, again for fixed values of $M$ and $r_{\text{h}}$. The total potential energy of an equilibrium dynamical model is an important dynamical quantity, because it sets the equipartition of the total energy budget due to the virial theorem \citep{2008gady.book.....B}, it is one of the ingredients to quantify the 3D concentration of dynamical systems \citep{1969ApJ...158L.139S}, and it sets the preferred length scale for Monte Carlo or N-body simulations \citep{1971Ap&SS..14..151H, 1979ApJ...234.1036C, 1986LNP...267..233H}. The solid line in this panel represents the analytical value from \citet{2019A&A...630A.113B},
\begin{multline}
W_{\text{tot}} = -\frac{GM^2}{r_{\text{h}}}\,\frac{1}{d^n}
\left[\frac{\Gamma(2n)}{\Gamma(3n)} \right.
\\
\left.+ \frac{\Gamma(5n)}{2n\,\Gamma^2(3n)}\,{}_2F_1(2n,5n;2n+1;-1)\right],
\end{multline}
and the {\tt{SpheCow}} values perfectly reproduce these analytical values. The total potential energy is also not a monotonic function of $n$: it first gradually increases, or rather becomes less negative, as $n$ grows. The maximum value is reached at $n=2.17$, after which it grows more negative again.

\section{Isotropic dynamical models}
\label{isotropic.sec}

The properties described in the previous section only depend on the spatial distribution of the matter. While the density is a crucial quantity and usually the first quantity by which dark matter haloes are described, it does not provide the full dynamical picture. A full characterisation of an equilibrium dynamical model is contained within the phase-space distribution function. For any spherical density profile $\rho(r)$, it is possible to generate many different distribution functions, depending on the orbital structure of the system \citep[e.g.][]{1986PhR...133..217D, 2008gady.book.....B, 2021isd..book.....C}. The simplest models are probably ergodic or isotropic dynamical models, that is, models in which the distribution function only depends on $\calE$, the binding energy per unit mass.

\begin{figure*}
\includegraphics[width=0.95\textwidth]{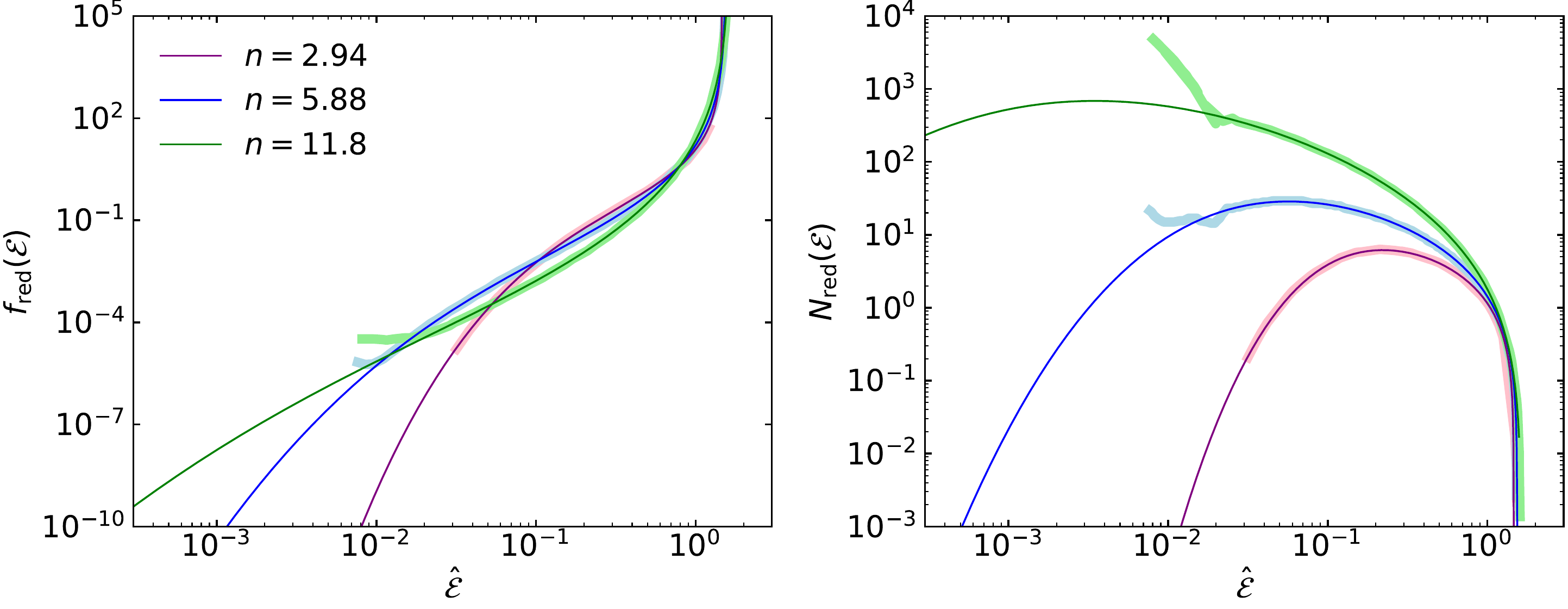}%
\caption{Comparison of the distribution function and differential energy distribution presented by \citet{2005MNRAS.358.1325C} (thick lines) with the corresponding values calculated with {\tt{SpheCow}} (thin lines). The different properties are represented in the dimensionless units convention adopted by \citet{2005MNRAS.358.1325C}, which differs from the representation adopted in the remainder of this paper.}
\label{Einasto-Cardone.fig}
\end{figure*}

\subsection{Velocity dispersions}

Panels (a) and (b) of Fig.~{\ref{Einasto-isotropic.fig}} show the intrinsic and line-of-sight velocity dispersion profiles. \citet{2005MNRAS.358.1325C} already showed velocity dispersion profiles for a number of Einasto models, but their study was limited to values $n>2.5$. Their dispersion profiles all have a strong central depression, but it is unclear from their Figure~5 (which is shown in a linear scale) how these profiles behave at small radii. We find that all Einasto models have a finite, non-zero central dispersion, in agreement with the fact that all Einasto models have a finite central density \citep{2002A&A...386..149B}. The solid line in the bottom panel in Fig.~{\ref{Einasto-central.fig}} shows how the central velocity dispersion depends on the Einasto index for fixed total mass and half-mass radius. It decreases smoothly and monotonically as $n$ increases.

All models with $n>\tfrac12$ have a central depression, implying that the dispersion profile first increases as a function of radius until a maximum value is reached, and subsequently decreases. The models with $n\leqslant\tfrac12$ have a monotonically decreasing dispersion profile over the entire radial range. This behaviour is also in agreement with the expectations \citep{2002A&A...386..149B}, as the density at small radii behaves as
\begin{equation}
\rho(r) \approx \rho_0 \left[1 - d\left(\frac{r}{r_{\text{h}}}\right)^{1/n}\right].
\end{equation}
At large radii, the intrinsic velocity dispersion profile is relatively shallow for large $n$ and it becomes gradually steeper as $n$ decreases. 

The line-of-sight velocity dispersion profiles behave in a very similar way as the intrinsic dispersion profiles. All Einasto models have a finite, non-zero central line-of-sight velocity dispersion, as indicated by the dashed line in the bottom panel of Fig.~{\ref{Einasto-central.fig}}. At fixed mass and half-mass radius, the central line-of-sight dispersion decreases smoothly and monotonically as $n$ decreases, but the decrease is gentler than for the intrinsic dispersion. Models with $n\leqslant\tfrac12$ have a monotonically decreasing line-of-sight velocity dispersion profile, whereas the profile for models with $n>\tfrac12$ have a central depression.

For any spherical density profile, {\tt{SpheCow}} automatically calculates the total kinetic energy of the corresponding isotropic dynamical model by numerically integrating the velocity dispersion profile,
\begin{equation}
T_{\text{tot}} = 6\pi \int_0^\infty \rho(r)\,\sigma^2(r)\,r^2\,\txd r.
\end{equation} 
For each of the Einasto models considered, we found that the kinetic energy $T_{\text{tot}}$ and the potential energy $W_{\text{tot}}$ relate as $W_{\text{tot}} + 2\,T_{\text{tot}} = 0$, as required by the virial theorem \citep{2008gady.book.....B}.

\subsection{Distribution function}
\label{dfiso.sec}

Panel (c) of Fig.~{\ref{Einasto-isotropic.fig}} shows the isotropic distribution function for the Einasto models. It is immediately clear that the shape of the distribution function varies systematically with the Einasto index. \citet{2005MNRAS.358.1325C} mention that the distribution function of Einasto models is well approximated by a single power law for a large range of values of the binding energy. We cannot confirm this, as we find a systematically changing slope of the distribution function as $n$ decreases. It is worth noting that \citet{2005MNRAS.358.1325C} mention that they have encountered technical problems in approaching the lower and upper ends of the range for the binding energy, and therefore they have computed the distribution function for their models only over a limited range. In the left panel of Fig.~{\ref{Einasto-Cardone.fig}} we directly compare the distribution functions as presented by \citet{2005MNRAS.358.1325C} with the {\tt{SpheCow}} calculations, for the same set of Einasto models. We converted the {\tt{SpheCow}} output to the dimensionless units convention adopted by them, which differs from the representation adopted in the remainder of this paper.\footnote{The original data from \citet{2005MNRAS.358.1325C} were retrieved using the freely available {\tt{WebPlotDigitizer}} software \citep{Rohatgi2020}.} The results agree very well over most of the binding energy range, but they clearly deviate at the lowest binding energies they considered. Over a limited range of binding energies, the distribution function of the Einasto models with $2.5\lesssim n \lesssim 10$ can be approximated by a single power law. This similarity breaks down when considering a larger range in binding energies, or when considering a larger range in Einasto parameters.

A crucial requirement for physical dynamical models is that the distribution function is positive over the entire phase space. For spherical models with a given density profile, it is always possible to formally calculate the isotropic distribution function $f(\calE)$ using the well-known Eddington formula \citep{2008gady.book.....B}. However, only when this distribution function is positive over the entire range of binding energies $0<\calE\leqslant\Psi_0$, the dynamical model is physically consistent and meaningful. This requirement is not always satisfied: examples of simple density profiles that cannot be supported by an isotropic velocity distribution include the uniform density sphere \citep{Zeldovich72, 1979PAZh....5...77O, 2022MNRAS.512.2266B}, S\'ersic models with S\'ersic index $m<\tfrac12$ \citep{2019A&A...626A.110B}, broken power-law models \citep{2020ApJ...892...62D, 2021MNRAS.503.2955B}, and double power-law models with sharp transitions \citep{1997MNRAS.287..525Z, 2021MNRAS.503.2955B}.

It turns out that not all Einasto models can be supported by an isotropic distribution function. For large values of $n$, the distribution function is a strongly increasing function of binding energy that is positive over the entire range of binding energies, and it diverges as $\calE$ approaches $\Psi_0$. As $n$ decreases, $f(\calE)$ still increases strongly for small $\calE$, but it starts to flatten at intermediate binding energies and it subsequently diverges again as $\calE$ approaches $\Psi_0$. For $n=\tfrac12$ the distribution function no longer diverges for $\calE\to\Psi_0$, but converges to a finite value. All models with $n<\tfrac12$ have a formal distribution function that increases as a function of $\calE$ for small $\calE$ until a maximum value is reached. It subsequently decreases and assumes negative values at the largest binding energy values (this is not visible in the plot as it uses logarithmic scaling). The particular $n=\tfrac12$ model, characterised by a Gaussian density profile, is the limiting model that still allows a positive isotropic distribution function. Any Einasto model with $n<\tfrac12$ can hence not support an isotropic orbital structure. 

This physical inconsistency obviously also accounts for the limiting case of the uniform density sphere, which corresponds to $n=0$. For this model, the formal isotropic distribution function can be calculated explicitly. It vanishes for $\calE < \calE_{\text{T}}$, it has an infinitely sharp peak at $\calE = \calE_{\text{T}}$, and it is negative for all binding energies $\calE_{\text{T}} < \calE < \Psi_0$, where 
\begin{gather}
\calE_{\text{T}} = \frac{3}{2}\,\Psi_0 = \frac{3}{2\sqrt[3]{2}}\,\frac{GM}{r_{\text{h}}}.
\end{gather}
This is obviously not the distribution function of physically consistent model. For more details we refer to \citet{2022MNRAS.512.2266B}.

\subsection{Differential energy distribution}

A final important dynamical quantity to consider is the differential energy distribution. It represents the distribution of the total mass as a function of binding energy and it has been argued to be the most fundamental partitioning of an equilibrium dynamical system \citep{1982MNRAS.200..951B, 2007LNP...729..297E, 2010ApJ...722..851H}. 

\citet{2005MNRAS.358.1325C} presented an intriguing result concerning the low binding-energy limit of the differential energy distribution of the Einasto models. They found that $N(\calE)$ converged to zero in the limit $\calE\to0$ for the smallest value of $n$ they considered, but that it appeared to diverge for the largest value of $n$. They argue that such a diverging differential energy distribution is possible because the mass density formally vanishes only at infinity, and it is always possible to find stars at larger and larger radii with lower and lower binding energies so that $N(\calE)$ may diverge in the limit $\calE\to0$. In the right panel of Fig.~{\ref{Einasto-Cardone.fig}} we overplot the differential energy distribution profiles presented by \citet{2005MNRAS.358.1325C} with the corresponding {\tt{SpheCow}} results. From this plot, it is clear that the differential energy distribution does not diverge for any Einasto model. The apparent divergence in the \citet{2005MNRAS.358.1325C} differential energy distribution is the combined result of numerical problems in their integrations at low binding energies and the limited range of binding energies considered.

In panel~(d) of Fig.~{\ref{Einasto-isotropic.fig}} we show the differential energy distribution for the family of Einasto models covering a wide range of $n$. In all cases, $N(\calE)$ is a function that disappears at both small and large binding energies with a peak somewhere in the middle. For large values of $n$, the differential energy distribution peaks for small binding energies, indicating that the majority of the stars or particles are orbiting on loosely bound orbits. As $n$ decreases, the Einasto models gradually become more centrally concentrated and the differential energy distribution becomes narrower and peaked at gradually larger binding energies. The Gaussian model $n=\tfrac12$ is the critical Einasto model that can still be supported by an isotropic distribution, and hence for which also the differential energy distribution is positive over the entire range of binding energies. For models with smaller values of $n$, the formal differential energy distribution is characterised by a strong peak at relatively large binding energies, but also by negative values at the largest binding energies. In the limit of the uniform density sphere, the differential energy distribution is zero for $\calE<\calE_{\text{T}}$, infinitely large for $\calE=\calE_{\text{T}}$, and negative for $\calE_{\text{T}} < \calE < \Psi_0$ \citep{2022MNRAS.512.2266B}.

As a sanity check on the accuracy of its results, {\tt{SpheCow}} integrates the differential energy distribution over the entire range of binding energies, which should yield the total mass, 
\begin{equation}
\int_0^{\Psi_0} N(\calE)\,\txd\calE = M.
\end{equation}
A second sanity check consists of calculating the total integrated binding energy \citep{2021A&A...653A.140B}, which should relate to the total kinetic and potential energy as 
\begin{equation}
\int_0^{\Psi_0} N(\calE)\,\calE\,\txd\calE \equiv B_{\text{tot}} = 3\,T_{\text{tot}} = -\tfrac32\,W_{\text{tot}}.
\end{equation}
All isotropic Einasto models, including the physically unacceptable models with $n<\tfrac12$,  pass these sanity checks.

\section{Osipkov-Merritt dynamical models}
\label{OM.sec}

Isotropy is just one of the possible dynamical structures for spherical models. A popular alternative option are the Osipkov-Merritt models \citep{1979PAZh....5...77O, 1985AJ.....90.1027M}, which are characterised by an anisotropy profile that is isotropic in the central regions and radially anisotropic at large radii. Osipkov-Merritt models have been explored for many different density profiles \citep[e.g.][]{1985MNRAS.214P..25M, 1985AJ.....90.1027M, 1995MNRAS.276.1131C, 1997A&A...321..724C, 2001MNRAS.321..155L, 2002A&A...393..485B}. This popularity has two reasons. Firstly, the dynamical structure of Osipkov-Merritt models can be explored using equations that are quite similar to the isotropic case. In particular, the distribution function can be calculated using an inversion algorithm that is similar to the Eddington equation in the isotropic case \citep[e.g.][]{2008gady.book.....B, 2021isd..book.....C}. Secondly, the idea of dynamical models with an increasing anisotropy profile are interesting from a physical point of view. Most numerical N-body simulations for dark matter haloes are roughly isotropic in the central regions and radially anisotropic at larger radii \citep{2001ApJ...563..483T, 2004MNRAS.352..535D, 2011MNRAS.415.3895L, 2012ApJ...752..141L, 2013MNRAS.434.1576W, 2016MNRAS.462..663B}. 

For any spherical density profile, {\tt{SpheCow}} generates a family of Osipkov-Merritt models and calculates the most important dynamical quantities. Osipkov-Merritt models are characterised by an additional free parameter $r_{\text{a}}$, known as the anisotropy radius. For $r\ll r_{\text{a}}$ the orbital structure is isotropic, for $r\gg r_{\text{a}}$ radial orbits are dominant. For $r_{\text{a}}\to\infty$ the Osipkov-Merritt models reduce to the isotropic models, for $r_{\text{a}}=0$ the model only contains radial orbits. 

In Fig.~{\ref{Einasto-OM.fig}} we show a number of dynamical properties of an Einasto model assuming an Osipkov-Merritt orbital structure for different values of the anisotropy radius $r_{\text{a}}$. We selected the Einasto model with $n=2$, but the analysis does not depend on this particular choice, and qualitatively the same results are obtained for any other value of $n$ (at least for $n>\tfrac12$).

\subsection{Velocity dispersions}

\begin{figure*}
\hspace*{2em}%
\includegraphics[width=0.9\textwidth]{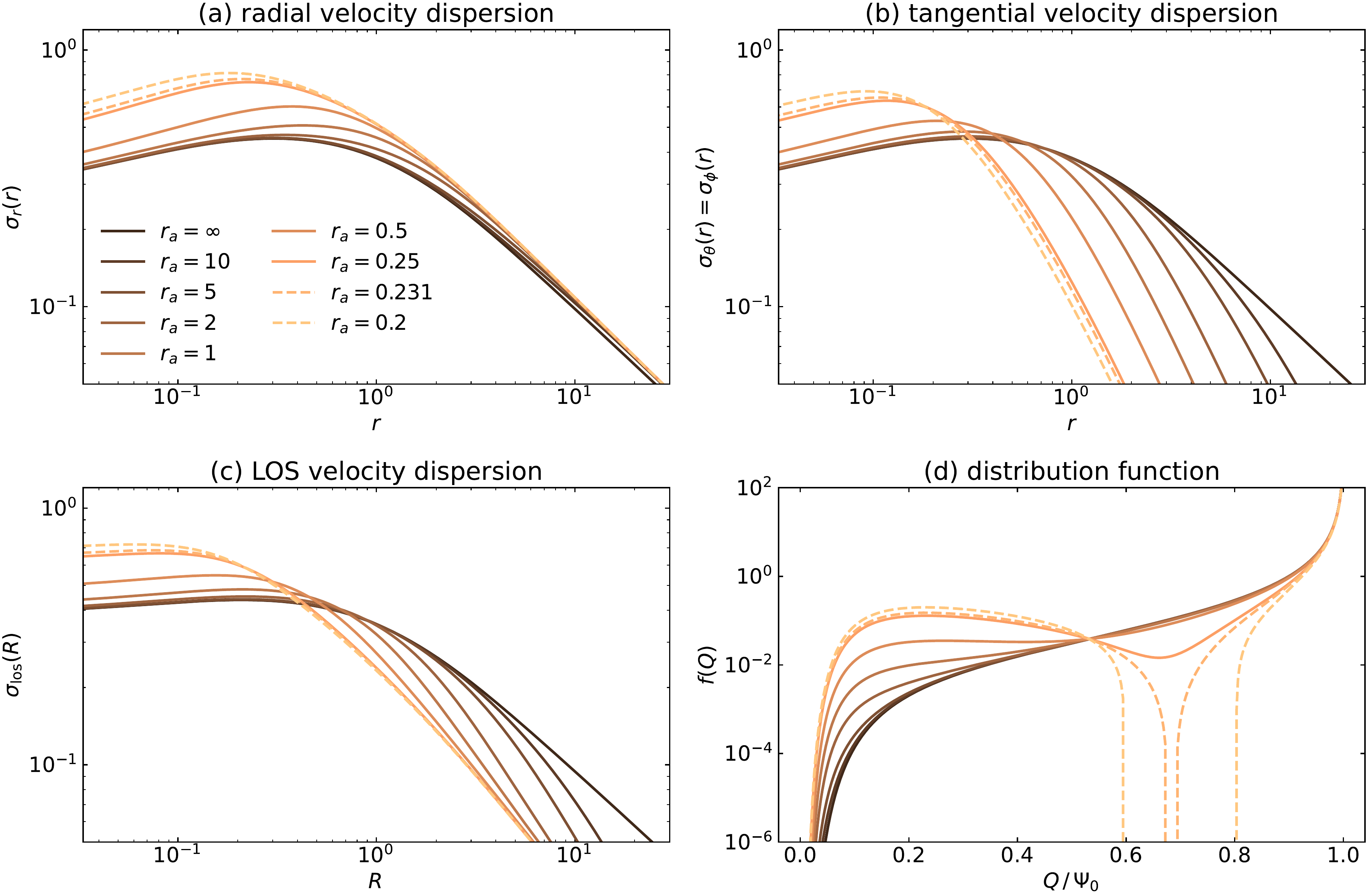}%
\caption{
Dynamical properties of the $n=2$ Einasto model with an Osipkov-Merritt anisotropic orbital structure. All lines correspond to an Einasto model with exactly the same density profile with $M=r_{\text{h}}=1$ and $n=2$, but with a different value for the anisotropy radius $r_{\text{a}}$, as indicated in panel (a). Solid lines are consistent dynamical models, dashed lines are unphysical dynamical models. The critical value for the anisotropy radius for this specific Einasto model is $(r_{\text{a}})_{\text{c}} = 0.2313$ (see Table~{\ref{Einasto-racrit.tab}}).}
\label{Einasto-OM.fig}
\end{figure*}

Panels (a) and (b) of Fig.~{\ref{Einasto-OM.fig}} show the radial and tangential velocity dispersion profiles. In the limit $r_{\text{a}}\to\infty$ the Osipkov-Merritt orbital structure turns into an isotropic dynamical structure, and the radial and tangential velocity dispersion profiles are identical to the velocity dispersion profile of the isotropic model presented in panel (a) of Fig.~{\ref{Einasto-isotropic.fig}}. For increasingly small $r_{\text{a}}$, the orbital structure changes from isotropic in the centre to radial at increasing smaller radii. This results in a radial velocity dispersion profile that increases with decreasing $r_{\text{a}}$ at all radii. The behaviour of the tangential velocity dispersion profile is slightly more complex: when $r_{\text{a}}$ decreases, the tangential velocity dispersion at large radii decreases strongly, as the outer regions become more and more dominated by radial orbits and devoid of circular-like orbits. At small radii, however, the tangential velocity dispersion increases in a similar way as the radial velocity dispersion, as the region inside the anisotropy radius remains roughly isotropic. In all cases, the radial and tangential velocity dispersion at the centre is finite and non-zero, as in the case of the isotropic model. 

The line-of-sight velocity dispersion profile, presented in panel (c), follows the same behaviour as a function of $r_{\text{a}}$ as the tangential velocity dispersion profile. At small projected radii, the component of the velocity ellipsoid that dominates along the line of sight is the radial component and the line-of-sight velocity dispersion increases with decreasing $r_{\text{a}}$. At large projected radii, the tangential component of the velocity ellipsoid dominates in the direction of the line of sight and the line-of-sight velocity dispersion is a decreasing function of decreasing $r_{\text{a}}$.

Also for the Osipkov-Merritt orbital structure, {\tt{SpheCow}} numerically calculates the total kinetic energy by integrating the velocity dispersion profiles,
\begin{equation}
T_{\text{tot}} = 2\pi \int_0^\infty \rho(r)\left[\sigma_r^2(r) + \sigma_\theta^2(r) + \sigma_\phi^2(r)\right] r^2\,\txd r.
\end{equation}
For any member of our family of Einasto models, we find that the total kinetic energy is independent of $r_{\text{a}}$, in agreement with the virial theorem.

\subsection{Distribution function}

Whether the velocity dispersion profiles discussed in the previous subsection are physically meaningful depends on whether the distribution function is positive over the entire phase space. Osipkov-Merritt models have a distribution function that is ellipsoidal in velocity space. More specifically, it can be written as $f(Q)$ with
\begin{equation}
Q = \calE - \frac{L^2}{2r_{\text{a}}^2},
\end{equation}
where $L$ is the angular momentum per unit mass. The distribution function $f(Q)$ for our set of $n=2$ Einasto models is shown in panel (d) of Fig.~{\ref{Einasto-OM.fig}}. 

For $r_{\text{a}}\to\infty$, $Q$ reduces to the binding energy $\calE$ and the Osipkov-Merritt distribution function reduces to the isotropic distribution function, which is a positive and monotonically increasing function of binding energy, as already demonstrated in the previous section. As $r_{\text{a}}$ decreases, the distribution function does not change in the large $Q$ limit ($Q\lesssim\Psi_0$), which corresponds to the isotropic central regions. In the small $Q$ regime, the distribution function does change with decreasing $r_{\text{a}}$: due to an increase in the number of weakly bound radial orbits at large radii, $f(Q)$ gradually increases with decreasing $r_{\text{a}}$. When $r_{\text{a}}$ becomes small enough, the distribution function is no longer a monotonically increasing function of $Q$, but shows a local minimum (see for example the $r_{\text{a}} = 0.25$ model in Fig.~{\ref{Einasto-OM.fig}}). When $r_{\text{a}}$ decreases below a critical value $(r_{\text{a}})_{\text{c}}$, the depression in the distribution function becomes so deep that it reaches negative values. At this point, the Osipkov-Merritt model is no longer physical. Any model with $r_{\text{a}}<(r_{\text{a}})_{\text{c}}$ is physically inconsistent. For the Einasto $n=2$ model presented in Fig.~{\ref{Einasto-OM.fig}}, the critical value is $(r_{\text{a}})_{\text{c}} = 0.2313$.

\begin{figure}
\includegraphics[width=0.9\columnwidth]{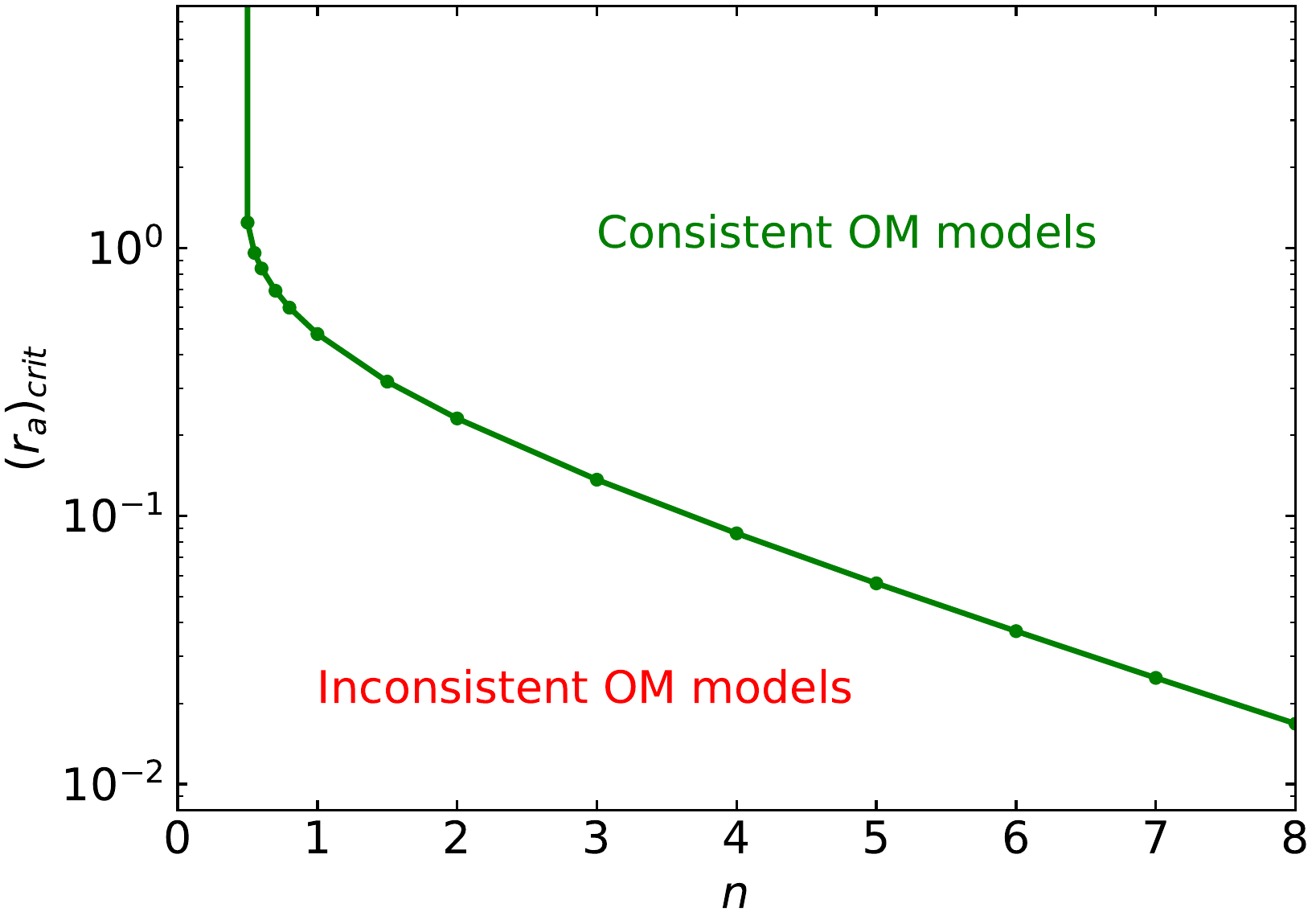}%
\caption{Critical anisotropy radius $(r_{\text{a}})_{\text{c}}$ as a function of the Einasto index $n$. Models with $r_{\text{a}} > (r_{\text{a}})_{\text{c}}$ have a positive Osipkov-Merritt distribution function, the models below or to the left of the green line are inconsistent.}
\label{Einasto-racrit.fig}
\end{figure}

\begin{table}
\caption{Critical anisotropy radius $(r_{\text{a}})_{\text{c}}$ for a number of Einasto models with index $n$.} 
\label{Einasto-racrit.tab}
\centering
\begin{tabular}{cc}
\hline\hline\\[-0.5em]
$n$ & $(r_{\text{a}})_{\text{c}}$ \\[0.5em]
\hline \\[-0.5em]
0.5 & 1.2453 \\
0.6 & 0.8391 \\
0.7 & 0.6931 \\
0.8 & 0.5993 \\
1 & 0.4778 \\
2 & 0.2313 \\
3 & 0.1367 \\
4 & 0.0862 \\
6 & 0.0372 \\
8 & 0.0168 \\
$\infty$ & 0 \\[0.5em] 
\hline\hline
\end{tabular}
\end{table}

The fact that a critical limiting value $(r_{\text{a}})_{\text{c}}$ exists is logical, given that a completely radial orbital distribution function can only be sustained by models with a density profile at least as steep as $r^{-2}$ at all radii \citep{1984ApJ...286...27R, 1992MNRAS.255..561C}. A similar behaviour for the distribution function for Osipkov-Merritt models as a function of $r_{\text{a}}$ has been noted for other models, such as the Plummer model  \citep{1985AJ.....90.1027M}, the $\gamma$-models \citep{1995MNRAS.276.1131C}, and the S\'ersic models \citep{1997A&A...321..724C}.

In Table~{\ref{Einasto-racrit.tab}} we list numerical values for $(r_{\text{a}})_{\text{c}}$ for a number of Einasto models and in Fig.~{\ref{Einasto-racrit.fig}} we graphically show the dependence of $(r_{\text{a}})_{\text{c}}$ on $n$. In Sec.~{\ref{dfiso.sec}} we showed that the Gaussian model $n=\tfrac12$ is the limiting model that can still be supported by an isotropic distribution function. Since any isotropic model with $n<\tfrac12$ is inconsistent, the same obviously accounts for an Osipkov-Merritt orbital structure, which is even more demanding in terms of consistency. For the limiting case, the Gaussian model, we find $(r_{\text{a}})_{\text{c}} = 1.2453$. So the Gaussian model $n=\tfrac12$ does not only support isotropic dynamical models, but also a range of anisotropic Osipkov-Merritt models, as long as the anisotropy radius is large enough. When $n$ increases, $(r_{\text{a}})_{\text{c}}$ decreases roughly exponentially, implying that Einasto models with larger Einasto indices can support more extreme radially anisotropic models. A similar systematic behaviour is found for the related set of S\'ersic models by \citet{1997A&A...321..724C}. In Sec.~{\ref{density.sec}} we discussed that the Einasto model reduces to a pure power-law model with a density profile proportional to $r^{-3}$ in the limit $n\to\infty$. This model can in principle support a purely radial orbital structure, so we formally find $\lim_{n\to\infty}(r_{\text{a}})_{\text{c}} = 0$.

\section{Discussion}
\label{Discussion.sec}

The Einasto model has become one of the most popular models to describe the density profile of dark matter haloes \citep[e.g.][]{2006AJ....132.2685M, 2008MNRAS.387..536G, 2008MNRAS.388....2H, 2008MNRAS.391.1685S, 2010MNRAS.402...21N, 2014MNRAS.439..300L, 2016MNRAS.457.4340K, 2020MNRAS.499.2426F, 2020Natur.585...39W}. This popularity demands a thorough investigation of the intrinsic and dynamical properties of this model. Up to now there have been relatively few comprehensive studies on the dynamical structure of the Einasto model, mainly because only a limited number of properties can be calculated analytically. 

In this paper we have used of the capabilities of the {\tt{SpheCow}} code \citep{2021A&A...652A..36B} to numerically explore the photometric and dynamical structure of the family of Einasto models, covering the entire range of Einasto indices. Our main results are the following:
\begin{itemize}
\item We present a new fitting formula for the dimensionless parameter $d$ that appears in the formal expression for the density profile of the Einasto model. This expression is valid for small $n$ and it complements the asymptotic expression derived by \citet{2012A&A...540A..70R} for large $n$. 
\item Not all Einasto models can be supported by an isotropic distribution function. For $n>\tfrac12$ the distribution function is an increasing function of binding energy that is positive over the entire range of binding energies and that diverges as $\calE\to\Psi_0$. For $n = \tfrac12$, the distribution function converges to a finite value in the high binding-energy limit. All models with $n<\tfrac12$ have a formal isotropic distribution function that becomes negative at the highest binding energies and hence cannot be supported by an isotropic orbital structure. The same accounts the limiting case $n\to0$, the uniform density sphere, as already found by previous studies \citep{Zeldovich72, 1979PAZh....5...77O, 2022MNRAS.512.2266B}.
\item The differential energy distribution for all isotropic Einasto models converges to zero in the low binding-energy limit. We can confirm that the intriguing apparent divergence reported by \citet{2005MNRAS.358.1325C} is the result of numerical problems in their integrations at low binding energies and the limited range of binding energies considered.
\item For each Einasto model with $n>\tfrac12$, a family of anisotropic dynamical models with a radial Osipkov-Merritt orbital structure can be considered. We find a critical value $(r_{\text{a}})_{\text{c}}$ for each value of $n$ that corresponds to the minimum anisotropy radius for the Osipkov-Merritt dynamical model to remain consistent: for smaller values of $r_{\text{a}}$ the formal distribution function reaches negative values. Since Einasto models with large $n$ are less centrally concentrated than models with small $n$, the critical anisotropy radius is a decreasing function of $n$.    
\end{itemize}
We double-checked the results of our {\tt{SpheCow}} calculations through comparisons with analytical results where available, by performing numerical integrations of the distribution function and the differential energy distribution, and by checking the general energy relations for dynamical systems \citep{2021A&A...653A.140B}. 

This paper demonstrates that studies of the structure and dynamics of models for galaxies and dark matter haloes should not be restricted to completely analytical models. There is only a fairly restricted set of models in which properties such as the potential, velocity dispersion, and distribution function can be calculated analytically. While these have their obvious benefits, it would be a shame to concentrate on these models alone only because they have analytical properties. The present study shows that numerical codes such as {\tt{SpheCow}} can help to open up the range of models that are thoroughly and systematically investigated. 

In relation to this point, it is interesting to note that the Einasto model is not the ultimate model proposed to describe the spherically averaged density profile of dark matter haloes. In the past two years alone, at least three extensions of the Einasto models have been proposed. 
\begin{itemize}
\item \citet{2020MNRAS.497.2393L} proposed a modification of the standard Einasto profile to account for the effects of baryonic feedback on the dark matter density profile. The presence of baryons and the corresponding galaxy evolution physics can have various effects on dark matter density profiles: the central dark matter density can be boosted as a result of baryons clustering at the centre of the halo, or it can be decreased due to stellar or active galactic nucleus energy feedback, or by dynamical friction from accretion events. The core-Einasto profile proposed by \citet{2020MNRAS.497.2393L} extends the standard Einasto model with one additional free parameter, a core radius. They demonstrate that this new model describes the spherically averaged density profiles of dark matter haloes from the FIRE-2 simulations \citep{2018MNRAS.480..800H} very well. In particular, they demonstrate that this new core-Einasto profile provides a superior fit to the density of the models than the core-NFW model by \citet{2012ApJ...759L..42P}, which has the same number of parameters. 
\item A second generalisation of the standard Einasto profile was proposed by \citet{2020MNRAS.499.2426F}. Using a set of zoom-in simulations \citep{2013ApJ...763...70W, 2015ApJ...810...21M}, they investigated the density profiles of the smooth components of dark matter haloes by excluding mass contained within sub-haloes. They found that the smooth halo density profiles differ substantially from the conventional halo density profile, and in particular found a more rapid decline at large radii. Their generalised Einasto model has the density profile of the Einasto model modified by an additional power-law factor. They find that this generalised Einasto profile provides a better fit to the density profiles of both Milky Way-mass and cluster-mass haloes than a standard Einasto or an NFW profile. 
\item Based on the observation that dark matter haloes are not steady-state objects, but contain both orbiting and infalling components, \citet{2022MNRAS.513..573D} dynamically split the particles in dark matter halo simulations into orbiting and infalling components and analysed their density separate profiles. In a follow-up paper, \citet{2022arXiv220503420D} proposed a generalisation of the Einasto model, dubbed the truncated Einasto profile, to describe the orbiting term. This generalisation contains five rather than three free parameters. When fixing two of the five parameters, the resulting three-parameter model on average fits the density profile of individual dark matter haloes better than the three-parameter Einasto profile.
\end{itemize}
Since {\tt{SpheCow}} only requires an analytical density profile or surface density profile as a starting point, these extensions to the Einasto model are in principle easy to investigate in more detail. We invite colleagues to use and extend the publicly available {\tt{SpheCow}} code for such investigations.

\begin{acknowledgements}
MB thanks Benedikt Diemer for a stimulating discussion and for his feedback on this manuscript. The anonymous referee is acknowledged for a prompt and constructive report.
\end{acknowledgements}

\bibliography{Einasto_bib}

\end{document}